\documentclass[useAMS,usenatbib]{mn2e}
\usepackage{journal_names}
\usepackage{graphicx,times}
\usepackage{amsmath}
\usepackage[T1]{fontenc}
\usepackage{aecompl}
\usepackage{subfig}
\usepackage{tabularx}
\usepackage{pdflscape}
\usepackage{rotating}
\usepackage{multirow}

\def\la{\mathrel{\hbox{\rlap{\hbox{\lower4pt\hbox{$\sim$}}}\hbox{$<$}}}}
\def\ga{\mathrel{\hbox{\rlap{\hbox{\lower4pt\hbox{$\sim$}}}\hbox{$>$}}}}

\newcommand{\HI}{\mbox{\normalsize H\thinspace\footnotesize I}}

\def\apjl   {{ApJL }}
\def\apjs   {{ApJS }}

\title[21 cm \HI\ line spectra of ZOA galaxies]{NIR Tully-Fisher in the Zone of Avoidance.   --  II. 21~cm \protect\mbox{\huge H\thinspace\Large I}-line spectra of southern ZOA galaxies.}

\author[K.~Said et al.]
{\parbox{\textwidth}{Khaled~Said$^{1,2,3}$\thanks{E-mail: khaled@ast.uct.ac.za},
\ Ren\'ee~C.~Kraan-Korteweg$^{1}$,
Lister~Staveley-Smith$^{2,3}$,
Wendy L. Williams$^{1,4,5}$,
T.~H.~Jarrett$^{1}$, and
Christopher M. Springob$^{2,3}$} \vspace{0.4cm}\\
\parbox{\textwidth}{$^{1}$Astrophysics, Cosmology and Gravity Centre (ACGC), Astronomy Department, 
University of Cape Town, Private Bag X3, Rondebosch, 7701, South Africa\\
$^{2}$International Centre for Radio Astronomy Research (ICRAR), M468, The University of Western Australia, 35 Stirling Highway, Crawley, WA 6009, Australia\\
$^{3}$ARC Centre of Excellence for All-sky Astrophysics (CAASTRO)\\
$^{4}$Leiden Observatory, Leiden University, PO Box 9513, NL-2300 RA Leiden, the Netherlands\\
$^{5}$Netherlands Institute for Radio Astronomy (ASTRON), PO Box 2, NL-7990 AA Dwingeloo, the Netherlands} }

\begin{document}

\date{Accepted 2016 January 11.  Received 2015 December 24; in original form 2015 October 29}

\pagerange{\pageref{firstpage}--\pageref{lastpage}} \pubyear{2015}

\maketitle

\label{firstpage}

\begin{abstract}
High-accuracy \HI\ profiles and linewidths are presented for inclined ($(b/a)^o  < 0.5$) spiral galaxies in the southern Zone of Avoidance (ZOA).  These galaxies define a sample for use in the determinations of peculiar velocities using the near-infrared Tully-Fisher (TF) relation. The sample is based on the 394 \HI-selected galaxies from the Parkes \HI\ Zone of Avoidance survey (HIZOA). Follow-up narrow-band Parkes \HI\ observations were obtained in 2010 and 2015 for 290 galaxies, while for the further 104 galaxies, sufficiently high signal-to-noise spectra were available from the original HIZOA data. All 394 spectra are reduced and parameterized in the same systematic way. Five different types of linewidth measurements were derived, and a Bayesian mixture model was used to derive conversion equations between these five widths.  Of the selected and measure galaxies, 342 have adequate signal-to-noise (S/N $\geq$ 5) for use in TF distance estimation. The average value of the signal-to-noise ratio of the sample is 14.7. We present the \HI\ parameters for these galaxies. The sample will allow a more accurate determination of the flow field in the southern ZOA which bisects dynamically important large-scale structures such as Puppis, the Great Attractor, and the Local Void.
\end{abstract}

\begin{keywords}
galaxies: spiral -- galaxies: distances and redshifts -- radio lines: galaxies -- infrared: galaxies -- cosmology: observations -- cosmology: large-scale structure of Universe
\end{keywords}

\section{Introduction}

Our understanding of the dynamics in the nearby Universe has increasingly been refined in the last decade thanks to the advance of larger and more systematic surveys of peculiar velocities using either  Tully-Fisher (TF) analyses such as Cosmic Flows (\citealt{2009AJ....138..323T,2011MNRAS.415.1935C,2012AN....333..436C}) and the 2MASS Tully-Fisher Survey (2MTF: \citealt{2008AJ....135.1738M,2013MNRAS.432.1178H,2014MNRAS.443.1044M,2014arXiv1409.0287H,2015arXiv151104849S}) or Fundamental Plane (FP) analyses such as 6dF (\citealt{2001MNRAS.321..277C,2012MNRAS.427..245M,2014MNRAS.443.1231C,2014MNRAS.445.2677S,2016MNRAS.455..386S}). 
The The Cosmic Flows project, for example, aims to measure distances for thousands of galaxies using new and archival 21-cm observations and $I$-band photometry. Longer wavelength photometry in the $K'$ band \citep{2009MNRAS.394.2022M} or using  {\sc Spitzer} \citep{2013ApJ...765...94S} or {\sc WISE} \citep{2014ApJ...792..129N} has been used to reduce the effect of dust extinction. The 2MTF project uses the infrared $J$, $H$, and $K_s$ bands in order to minimize the effect of the foreground extinction. However, a non-negligible portion of the sky remains excluded because the 2MTF selection also depends on the optical 2MASS Redshift Survey (2MRS; \citealt{2012ApJS..199...26H}). One of the major limitations therefore remains the lack of data in the so-called Zone of Avoidance (ZOA) around the Galactic Plane (e.g. \citealt{2000A&ARv..10..211K,2008MNRAS.386.2221L}). This region is known to bisect major parts of dynamically important nearby large-scale structures such as the Perseus-Pisces Supercluster (PPS; \citealt{1980MNRAS.193..353E,1982AJ.....87.1355G,1984A&A...136..178F,1987A&A...184...43H}), the Great Attractor (GA; \citealt{1988ApJ...326...19L,1999A&A...352...39W}) and the Local Void (LV; \citealt{1987ang..book.....T,2008glv..book...13K}).
As a result, a definitive analysis remains to be completed of fundamental questions such as whether the bulk motion of the local Universe is consistent with theoretical expectations, and which structures are  responsible for the bulk motion. Large forthcoming surveys will probe the local Universe to greater and greater accuracy. However, the uncertainties due to missing data in the ZOA need to be resolved by direct measurements in the region, if feasible.

In view of this situation, we are extending peculiar velocity surveys closer to the Galactic Plane. At high levels of dust extinction and stellar density, observations at the long wavelength of the 21 cm line of \HI\ and in the near-infrared (NIR) are little affected. An extension of NIR TF-based distances to lower latitudes based on deep 21cm observations and deep, well-resolved NIR imaging is therefore possible.

We are pursuing this TF extension as follows. Firstly, we use existing systematic \HI\ surveys, for which we obtain deep and well-resolved near-infrared photometry in order to select a sample suitable for a TF study -- with the subsequent intention of deeper or better resolved data where necessary. Secondly, we use existing NIR surveys to select galaxies. The Two Micron All-Sky Survey (2MASS; \citealt{2006AJ....131.1163S}) is extremely well-suited for this approach. 2MTF (\citealt{2008AJ....135.1738M}) already provides a complete Tully-Fisher analysis (i.e. distances and peculiar velocities) of all bright inclined spirals in the 2MRS (\citealt{2012ApJS..199...26H}). However, they exclude the inner ZOA  ($|b|<5^\circ$; $|b|<8^\circ$ for $|l|<30^\circ$).  Therefore, in 2009, we started an \HI\ redshift follow-up for all 2MASX bright galaxies without redshift data in or close to the ZOA, primarily using the Nan\c{c}ay Radio Telescope (NRT; \citealt{2014arXiv1412.5324R}; Kraan-Korteweg et al. in prep.), but later also the Parkes Telescope for a small remaining sample of very southern galaxies (Said et al. in prep.). This complementary approach is particularly useful, because no systematic \HI\ survey exists for the northern ZOA, while the 2MASX galaxy properties are not biased in any sense, being hardly affected by either extinction or high stellar densities in that part of the ZOA.

In preparation for the forthcoming flow field analysis, we have first derived a NIR TF relation that is optimized for areas of high extinction. Although the NIR parameters of galaxies are much less affected compared to optical data ($A_{\rm K} \sim 0.1 A_{\rm B}$) it is not entirely negligible, and total magnitudes will not be as reliable compared to high latitude regions. However, isophotal magnitudes are much more stable in regions affected by extinction. \citealt{2015MNRAS.447.1618S} (Paper I) therefore derived a new calibration of the NIR TF relation based on isophotal magnitudes. The relation is based on the same calibrator sample as the one presented in \cite{2008AJ....135.1738M} and follows their precepts closely. The isophotal TF relation was found to have a scatter of the same order as the total magnitude relation. A pilot project for an incomplete subset of the ZOA galaxy sample demonstrated the feasibility of extending flow fields deeper into the ZOA  \citep{SAIP_KS}. 

This paper is the first in a series of papers presenting data obtained in a systematic way for future applications of the TF relation in the ZOA. It focuses on the first approach, where we start out with an \HI-selected sample. It is based on the systematic \HI\ ZOA surveys pursued with the Parkes telescope, specifically designed to unveil the large-scale structures of galaxies in the nearby Universe across the most obscured part of the mostly southern ZOA (Staveley-Smith et al. 2015), i.e. in areas were most other surveys fail. The integration time of the HIZOA surveys are a factor of five longer compared to the \HI\ Parkes All-Sky Survey (HIPASS) \citep{2004MNRAS.350.1195M}. These \HI-selected galaxies are not biased with respect to the Galactic foreground dust layer, as shown in Staveley-Smith et al. (2015).

Three blind systematic deep \HI\ ZOA surveys were conducted with the Multibeam Receiver on the 64-m Parkes Radio Telescope, i.e. the main southern \HI\ Parkes Deep Zone of Avoidance Survey (HIZOA-S; Staveley-Smith et al., 2015), the Northern Extension (HIZOA-N) \citep{2005AJ....129..220D} and the Galactic Bulge (GB) \citep{2008glv..book...13K}. These three surveys show many new structures and also reveal the missing connections in the conspicuous features that cross the ZOA, such as the GA, the Puppis region and the LV for the first time. This highlights the importance of the ZOA in the understanding of the cosmic web. Based on these surveys, \HI\ spectra for over a thousand galaxies in the southern ZOA are now available.

Accordingly, follow-up pointing deep NIR observations for all galaxies in the three HIZOA surveys were conducted between 2006 and 2013 with the Japanese InfraRed Survey Facility (IRSF), a 1.4 m telescope situated at the South African Astronomical Observatory site in Sutherland. Data from 2006 to 2010 were published in \cite{2014MNRAS.443...41W}. In a companion paper, Said et al. (in prep.), we will present high-quality NIR $J$, $H$, and $K_s$-band observations for the completed HIZOA surveys, and discuss completeness and reliability of the NIR catalog for application in a TF-survey. The NIR data has already been used extensively for the sample selection for the present TF analysis with regard to unambiguous counterpart identification, availability of high accuracy photometry, and the required inclination limits.

The determination of the \HI\ parameters also required further attention. We need highly resolved \HI-line profiles, with high signal-to-noise and homogeneous analysis. 104 spectra in the ZOA TF sample already meet our requirements (S/N > 5, very well resolved \HI\ profile). Based on the newly-derived NIR imaging, an additional 290 needed to be re-observed to obtain higher resolution HI profiles. This was achieved by conducting 21-cm narrow-band follow-up observations with the Multibeam Receiver on the 64-m Parkes Radio Telescope.  

In this paper we present the newly obtained 21 cm \HI\ line spectra. We discuss the observations and data reduction procedures in Section 2. In Section 3 we describe the algorithm with which we derive the \HI\ parameters and associated errors, both for the new observations and the re-measurements of the HIZOA linewidths, the final compilation, the characteristics of the data set and comparison with the published \HI\ data. We summarize our results in Section 4. 


\section{Observations and Data Reduction}

\subsection{Sample Selection}

\begin{figure*}
\begin{center}
\includegraphics[scale=0.65,angle=270]{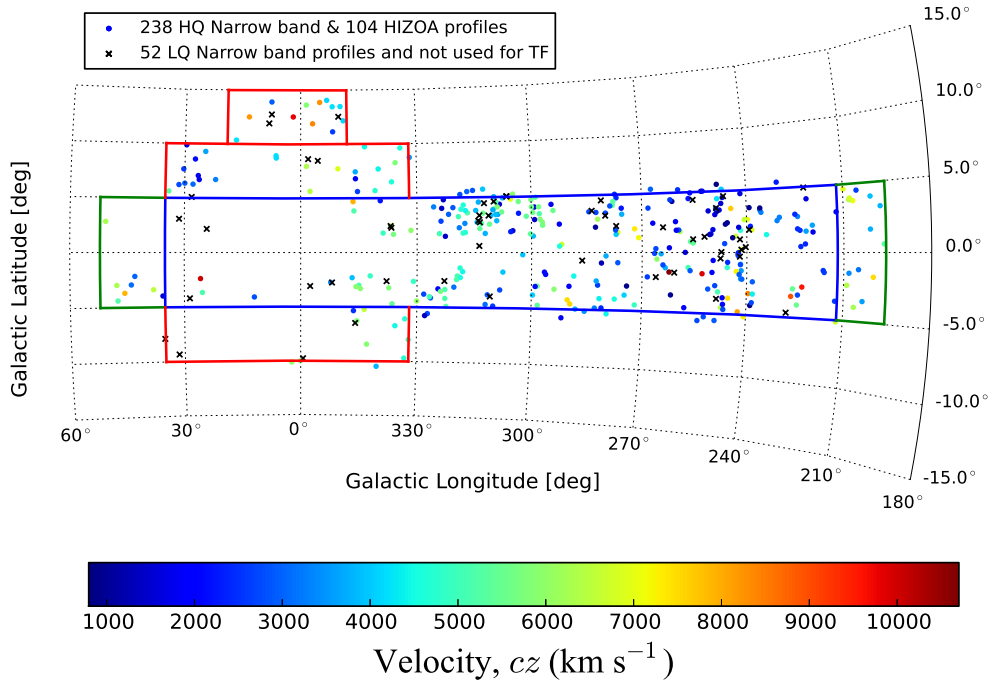}
\caption{The distribution of the 394 inclined spiral galaxies in the TF sample with accurate NIR photometry shown in a zoomed-in Aitoff projection: High Quality (HQ) narrow-band observations of 238 galaxies and 104 additional HIZOA galaxies with HQ \protect\HI\ profiles are shown as circles color coded by their velocity $cz$; Low Quality (LQ) narrow-band observations of the 52 galaxies discarded from the TF are shown as black crosses. The boundaries of the three HIZOA survey areas are plotted in blue (HIZOA-S), green (HIZOA-N) and red (GB). Note the wall-like structure of the Norma super-cluster at $290^\circ < l < 340^\circ$ and $cz \sim$ 4800 km s$^{-1}$.}
\label{All_HI_Par_compined_lb}
\end{center}
\end{figure*}

HIZOA is a blind \HI\ survey of the southern ZOA ($|b| < 5^\circ$; Dec $< +15^\circ$), conducted with the 64-m Parkes Radio Telescope\footnote{The Parkes telescope is part of the Australia Telescope National Facility which is funded by the Commonwealth of Australia for operation as a National Facility managed by CSIRO.} in three parts: (1) the main part (HIZOA-S) covers $|b| < 5^\circ; 212^\circ \le l \le 36^\circ$ (Staveley-Smith et al., 2015), (2) the Northern Extension (HIZOA-N) covers the northern regions visible from Parkes ($|b| < 5^\circ; 36^\circ < l < 52^\circ$ and $196^\circ < l < 212^\circ$;  \citealt{2005AJ....129..220D}), and (3) the Galactic Bulge (GB) extension, which goes to higher latitudes around the GB region ($5^\circ < |b| < 10^\circ; 332^\circ < l < 36^\circ$ and $10^\circ < b < 15^\circ; 348^\circ < l < 20^\circ$; \citealt{2008glv..book...13K}). These surveys cover the most obscured southern ZOA visible from Parkes out to 12000 km/s, and result in more than a thousand detected galaxies with an  rms~$\sim 6$ mJy/beam.


Our \HI\ sample selection is based on a cross-identification of the HIZOA galaxies with the high-quality NIR IRSF images. We select only galaxies with secure counterparts and good photometry (error in isophotal magnitude less than 0.1 mag at the highest extinction level). After applying additional restrictions on inclination, we use galaxies with $b/a<0.7$ (see also, Fig. 6 in \citealt{2015MNRAS.447.1618S}). All the selected HIZOA profiles were visually inspected. Galaxies with good profiles were excluded from the re-observation list. Only galaxies with profiles of insufficient S/N ratio for TF work were included in the narrow-band observing list. The final list for re-observations contains 290 galaxies. An additional 104 \HI\ profiles from the three HIZOA surveys were deemed sufficiently reliable for TF analysis and measured directly from data in the HIZOA archive.

\subsection{Data Acquisition and Reduction}
Data were collected in 2010 and 2015 using the 21 cm Multibeam receiver (MB; \citealt{1996PASA...13..243S}) on the 64-m Parkes Radio Telescope in narrow-band mode, using only the high-efficiency 7 inner beams. Beam-switching mode allows one beam ON the source and the other six OFF source, which reduces the noise by a factor of 2. Each target was observed for at least 35 minutes of ON-source integration time, with the 8 MHz bandwidth split into 1024 channels. This results in a velocity resolution of 1.6 km s$^{-1}$.

Preliminary processing of the data was done in the real time. Each galaxy was observed for at least 35 minutes to get improved sensitivity over the original HIZOA data. To avoid L3 beacon of the Global Positioning System (GPS), the integration time for sources near \textit{cz}= 8300 km s$^{-1}$ was reduced to 2 $\times$ 17.5 minutes. Based on an assessment of the preliminary 35 minutes integration results, some galaxies were observed for another 17.5 or 35 minutes. We used the package {\bf LIVEDATA}\footnote{This software is available in the ATNF package of AIPS++} \citep{2001MNRAS.322..486B} to correct for the bandpass and Doppler effects. Estimation of the bandpass is done by using the MEAN estimator. All spectra were converted to the Solar System barycentre. The corrected-spectra were gridded with {\bf GRIDZILLA}\footnote{This software is available in the ATNF package of AIPS++} \citep{2001MNRAS.322..486B}.

\section{\protect\HI\ Results}
As mentioned above, the aim of this paper is to provide \HI\ parameters measured in a uniform way from high signal-to-noise (S/N) ratio and High Quality (HQ) profiles to be used in the forthcoming NIR TF analysis of galaxies in the ZOA. The required S/N threshold of depends on the shape of the \HI\ profile, but S/N $\simeq$ 5 appears to be minimum requirement for our algorithm (Donley et al. 2015). In this section we describe how we obtain the \HI\ parameters required for the 394 galaxies selected for the TF ZOA relation. 

Figure \ref{All_HI_Par_compined_lb} shows the distribution of all 394 galaxies in the sample. The newly-observed HQ narrow-band observations of 238 galaxies and additional 104 HQ HIZOA profiles derived directly from the HIZOA surveys are shown as circles color-coded by their velocity $cz$. For 52 of the re-observed galaxies, the resulting profile or signal to noise ratios was not deemed of sufficient quality for the ZOA TF analysis. These 52 (LQ) narrow-band observations are shown as black crosses. In this figure, the wall-like structure of the Norma super-cluster seems to dominate the distribution at $290^\circ < l < 340^\circ$ and $cz \sim$ 4800 km s$^{-1}$ \citep{1996Natur.379..519K,1999A&A...352...39W,2008MNRAS.383..445W,2014MNRAS.439.3666M}. See also the wedge plot (Fig. 18) in Staveley-Smith et al. (2015).

\begin{figure}
\begin{center}
\includegraphics[scale=0.4,angle=270]{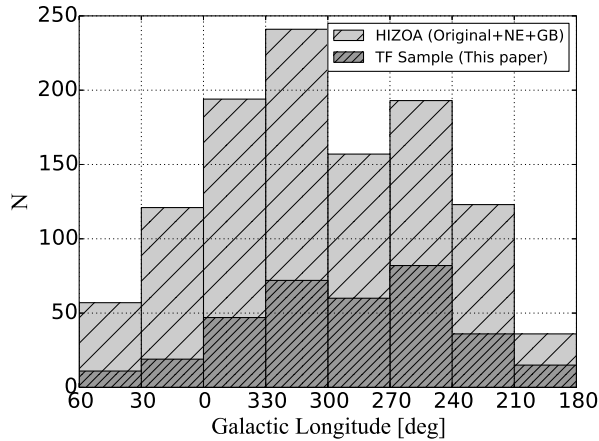}
\caption{The distribution of both HIZOA (N+S+GB) and the NIR TF ZOA sample as a function of Galactic longitude. The final TF sample is representative of the three input HIZOA surveys. The two peaks around $240^\circ < l < 270^\circ$ and $300^\circ < l < 330^\circ$ are due to Puppis and GA respectively. The drop in both histograms around the center of the Milky Way is due to existence of the LV. This drop is more prominent in the TF sample because of the dependence on accurate NIR photometry.}
\label{hist_hizoa_TF}
\end{center}
\end{figure}

Figure \ref{hist_hizoa_TF} presents a quantitative comparison between the three HIZOA surveys and the TF sample presented here.
The distribution of both the HIZOA sources and TF sample are shown as a function of Galactic longitude in Fig.~\ref{hist_hizoa_TF}. It shows that the TF sample is representative of the distribution of the parent HIZOA sample. The two histogram peaks around $240^\circ < l < 270^\circ$ and $300^\circ < l < 330^\circ$ in the HIZOA distribution are due to the Puppis and GA regions respectively. Both are well represented in the TF sample. The drop in both histograms towards the center of the Milky Way is due to existence of the LV. This drop is more prominent in the TF sample than the HIZOA survey because of the dependence of the TF sample on HQ NIR observations (Said et al. in prep.). In the Galactic Bulge region, not only is the stellar density high but the region is also dominated by the LV, and the sample contains a relatively larger proportion of smallish blueish dwarfs (Kraan-Korteweg et al. 2008). 

\subsection{\protect\HI\ Parametrization}

\subsubsection{The total flux and errors}
We follow Staveley-Smith L. et al. (2015) in measuring the total flux of the \HI\ using the function {\bf MBSPECT} within the {\bf MIRIAD} package \citep{1995ASPC...77..433S}. After applying three-channel Hanning smoothing, which results in a velocity resolution of 3.3 km s$^{-1}$ for our spectra, a low-order polynomial was fitted and subtracted from this spectrum. The integrated line flux was then measured from this smoothed and baseline-subtracted profile by integrating across the channels containing the emission line. The error was calculated using the formula given in \cite{2004AJ....128...16K}:

\begin{equation}
\sigma(F_{HI}) = 4(SN)^{-1}(S_{peak}F_{HI}\delta_v)^{1/2}
\end{equation}
where $(SN)$ is the ratio of the peak flux density, $S_{peak}$, and its uncertainty, and $\delta_v$ = 3.3 and 27 km s$^{-1}$ are the velocity resolution of narrow-band and HIZOA observations respectively, after applying three-channel Hanning smoothing. 
\subsubsection{Systemic velocities and velocity widths}
For the determination of the TF distances and peculiar velocities, the most important parameters from the \HI\ profile are the systemic velocity and linewidth. The measurement of the \HI\ linewidth and its associated uncertainty has received a lot of attention in the last few decades. \cite{1997AJ....113...22G} found the error in the \HI\ linewidth to be the dominant source for the scatter in the TF relation. However, \cite{2000ApJ...533..744T} state that this does not hold if the error is less than 20 km s$^{-1}$. Subsequently, different algorithms have been introduced to measure the linewidth with high precision. Some authors adopt $W_{F50}$, the linewidth at 50\% of the peak flux-rms ($f_p-$rms) measured with a polynomial fit to both sides of the profile, to be their first preference \citep{1999AJ....117.2039H,2005ApJS..160..149S}. \cite{2009AJ....138.1938C}, on the other hand, compared the $W_{20}$, the linewidth at 20\% of the peak flux of a single peak, as used in the early work \citep{1977A&A....54..661T,1988ApJ...330..579P} with the different linewidths from \cite{2005ApJS..160..149S} for an optimized choice  of linewidth. They found a better correlation of $W_{20}$ with $W_{M50}$, the linewidth at 50\% of the mean flux, than with $W_{F50}$. They therefore developed an algorithm to measure the $W_{m50}$ based on the mean flux instead of the peak level. This linewidth measurement, $W_{m50}$, has been used to update the Pre-Digital \HI\ catalog of the Extragalactic Distance Database \citep{2011MNRAS.414.2005C,2015MNRAS.447.1531C}.\\

In this paper, we measure the systemic velocities and velocity widths using a modified version of the {\bf GBTIDL}\footnote{GBTIDL is an interactive package for reduction and analysis of spectral line data taken with the Robert C. Byrd Green Bank Telescope (GBT) using IDL} function to find the Area, Width, and Velocity of a galaxy profile ({\bf AWV})\footnote{This code adapted from code in use at Arecibo. This version was originally from Karen O'Neil and modified by adding $W_{F50}$ option by Karen L. Masters} (see also, \citealt{2013MNRAS.432.1178H,2014MNRAS.443.1044M}). This allows us to measure the linewidth values using five different methods: 
\begin{enumerate}
\item $W_{P20}$: the linewidth at 20\% of the peak flux$-$rms,\\

\item $W_{M50}$: the linewidth at 50\% of the mean flux,\\

\item $W_{P50}$: the linewidth at 50\% of the peak flux$-$rms,\\

\item $W_{F50}$: the linewidth at 50\% of the peak flux$-$rms, measured with a polynomial fit to both sides of the profile,\\

\item $W_{2P50}$, the linewidth at 50\% of the peak flux$-$rms measured at each of the two peaks.
\end{enumerate}

The errors for the systemic velocities and velocity widths were calculated by conducting a set of simulations for each profile (see also, \citealt{2005AJ....129..220D,2013MNRAS.432.1178H}). This was achieved in three steps:

Step 1: The Savitzky-Golay smoothing filter \citep{1992nrfa.book.....P} was used to smooth each galaxy profile. A smoothing width of seventeen velocity channels was used for the new narrow-band observations, while eight velocity channels were used for HIZOA profiles (e.g., \citealt{2005AJ....129..220D,2013MNRAS.432.1178H}).

Step 2: Fifty simulated galaxy profiles were created for each galaxy by applying Poisson noise to the smoothed profile. The rms of the Poisson noise was adjusted to the rms of the original galaxy profile.

Step 3: A modified version of the {\bf GBTIDL} function {\bf AWV} was used again to measure all the parameters using the simulated spectra. The median offset between the 50 simulated spectra and the original spectrum was adopted as the error.

For 17 galaxies, the {\bf GBTIDL} function {\bf AWV} failed to determine the $W_{F50}$ from the simulated spectra. For these galaxies, we used the formula proposed by \cite{2014MNRAS.443.1044M}:
\begin{equation}
\epsilon_{WF50} = \sqrt{(rms/a_l)^{2}+(rms/a_r)^{2}}
\end{equation}
where $a_l$ and $a_r$ are the slopes of the fit on the left and right edge of the profile. This error was converted to the Monte Carlo simulation error using
\begin{equation}
\epsilon_{MC} = 0.815+0.405\epsilon_{WF50}
\end{equation}
also proposed by \cite{2014MNRAS.443.1044M}. 

\begin{figure*}
\def\tabularxcolumn#1{m{#1}}
\begin{tabularx}{\linewidth}{@{}cXX@{}}
\begin{tabular}{ccc}
\subfloat[]{\includegraphics[scale=0.4,angle=270]{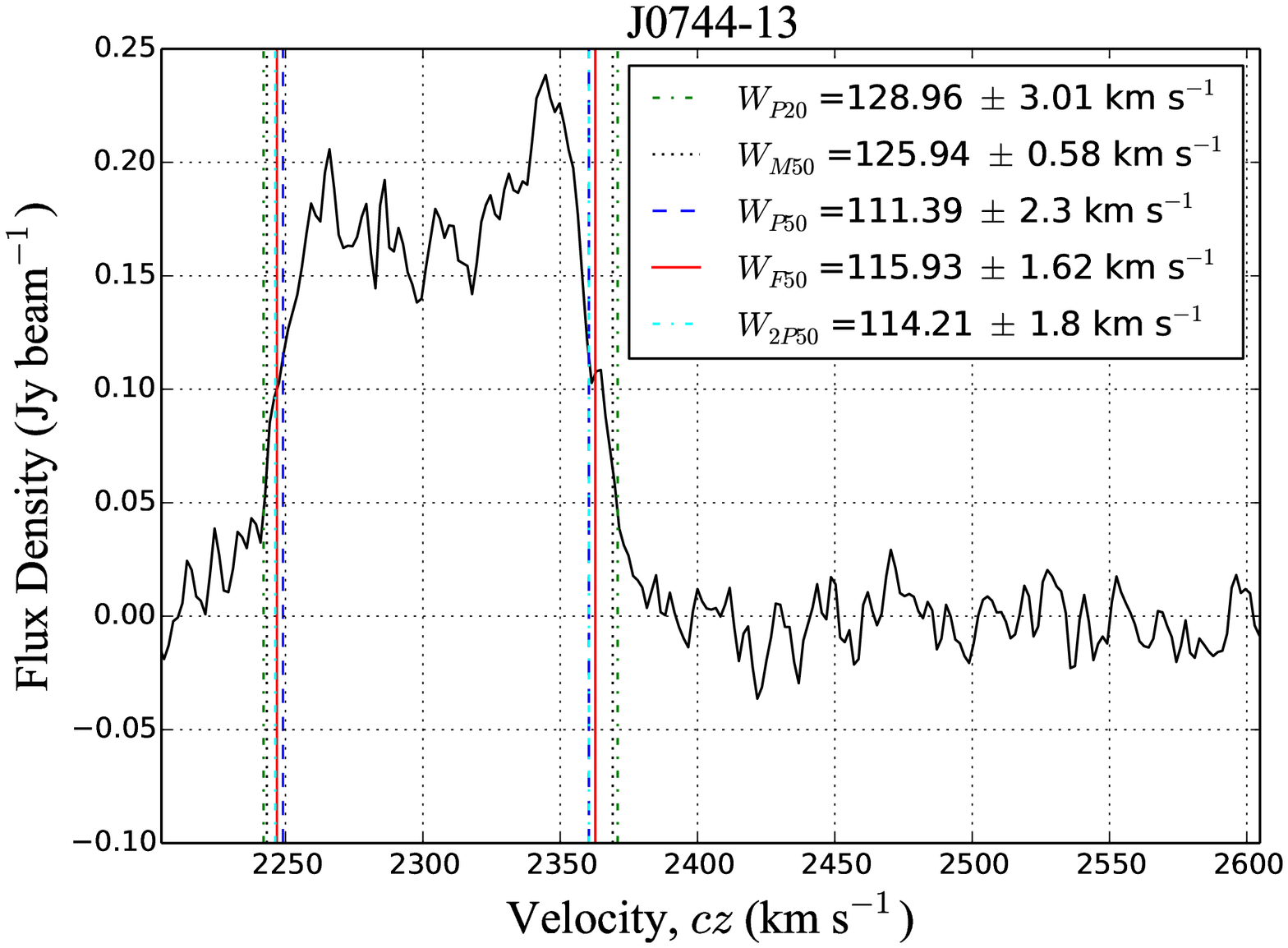}} 
   & \subfloat[]{\includegraphics[scale=0.4,angle=270]{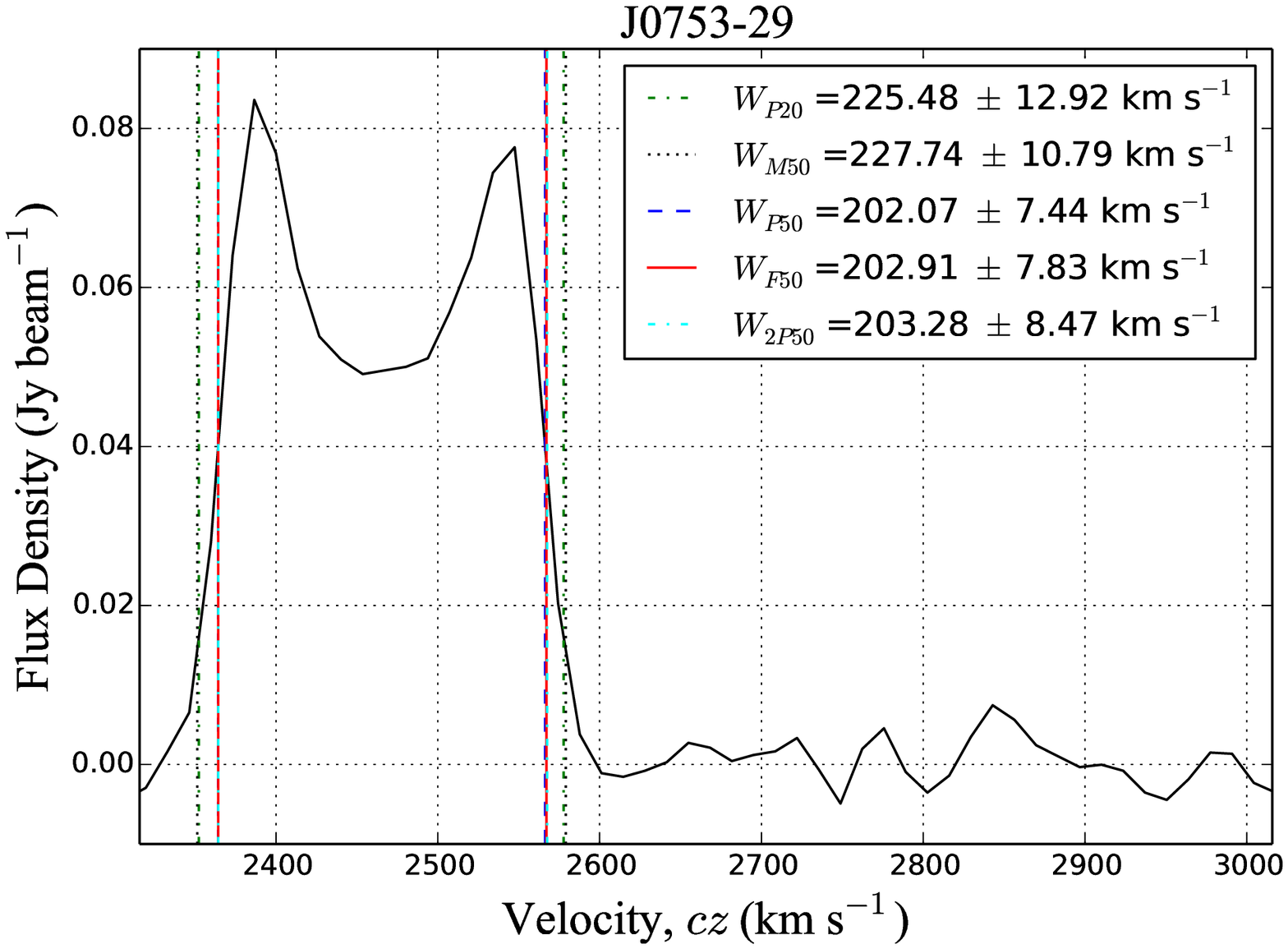}}
\end{tabular}

\end{tabularx}

\caption{Two examples of \protect\HI\ profiles in our sample measured with five different linewidth algorithms: (a) new narrow-band observations with velocity resolution of 3.3 km s$^{-1}$; (b) existing HIZOA data of a good \protect\HI\ profile, with a velocity resolution of 27 km s$^{-1}$. The solid red lines represents our preferred linewidth measurement $W_{F50}$. Figures for the entire sample are available on-line.}
\label{narrowandhizoa}
\end{figure*}

Figure \ref{narrowandhizoa}
shows two examples of \HI\ profiles in our sample measured with five different linewidth fitting algorithms. The profile on the left presents one of the newly observed galaxies with the narrow-band. The right profile is based on the HIZOA data but re-processed with the same method for the narrow-band observations. Figure \ref{narrowandhizoa} reveals good agreement within the uncertainty of $W_{P20}$, and $W_{M50}$ and also between $W_{F50}$ and $W_{2P50}$. 

\begin{figure*}
\begin{center}
\includegraphics[scale=0.58,angle=270]{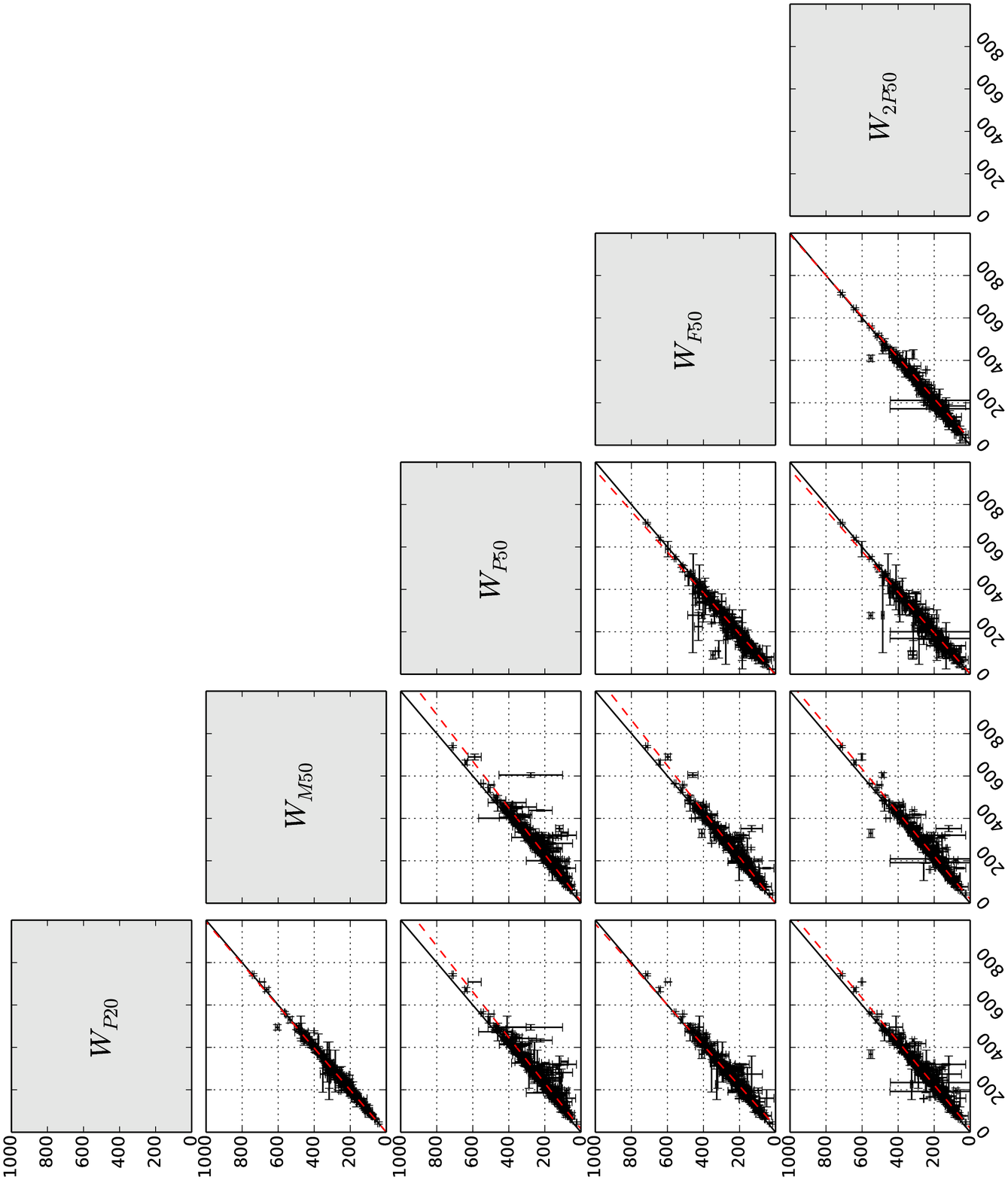}
\caption{A matrix of pairwise plots of the five linewidth measurements. The solid line shows the one-to-one relation, while the dashed red line is the best fit model to the data. Almost perfect correlation was found between $W_{P20}$ and $W_{M50}$ also between $W_{F50}$ and $W_{2P50}$ (within the uncertainty). A large offset was found between the two most used linewidths in the TF analysis $W_{M50}$ and $W_{F50}$.}
\label{compall}
\end{center}
\end{figure*}

To expand on this statement, we plot in Fig. \ref{compall} the five linewidth measurements against each other and determine their respective correlations.
The strongest correlation was found between $W_{P20}$ and $W_{M50}$  which confirms the results of \cite{2009AJ....138.1938C}. We also found good agreement between  $W_{F50}$ and $W_{2P50}$. In contrast, a significant offset was found between $W_{M50}$ and $W_{F50}$ which also agrees with what was found by \cite{2009AJ....138.1938C}. 

To quantify the correlations between the five linewidth measurements, we used the Pearson's sample correlation coefficient 
\begin{equation}
r = \frac{\sum_{i=1}^{n}(x_i-\bar{x})(y_i-\bar{y})}{\sqrt{\sum_{i=1}^{n}(x_i-\bar{x})^2}\sqrt{\sum_{i=1}^{n}(y_i-\bar{y})^2}}
\end{equation}
with $-1\leq r \leq1$, where $r\sim-1$ indicates a perfect anti-correlation and for uncorrelated parameters $r\sim0$, while $r\sim1$ means a strong positive correlation. Table \ref{PearsonC} shows the resulting Pearson's sample correlation coefficient correlation matrix.
\begin{table}
\begin{center}
\caption[The Pearson's sample correlation coefficient]{The Pearson's sample correlation coefficient $r$. The correlation coefficient clearly identifies $W_{F50}$ as the most stable linewidth with average correlation of $\bar{r}_{WF50}=0.975$ with the four other linewidths and therefore the optimal choice for TF analysis.}
\begin{tabular}{l c c c c c}
\hline
\hline
linewidth type  &   $W_{P20}$ & $W_{M50}$ & $W_{P50}$ & $W_{F50}$ & $W_{2P50}$\\
\hline
$W_{P20}$ &   1.000 &   &   &    &  \\
$W_{M50}$ &   0.993 &  1.000 &   &    &  \\
$W_{P50}$ &   0.947 &  0.932 &  1.000 &    &  \\
$W_{F50}$ &   0.978 &  0.974 &  0.953 &  1.000  &  \\
$W_{2P50}$&   0.973 &  0.970 &  0.956 &  0.993  &  1.000\\
\hline
\end{tabular}
\label{PearsonC}
\end{center}
\end{table} 
The best two correlations in Table \ref{PearsonC} are between $W_{P20}$ and $W_{M50}$ and between $W_{F50}$ and $W_{2P50}$ where $r=0.993$ for both of them. The least correlated linewidths are $W_{M50}$ and $W_{P50}$ where $r=0.932$. Based on the values of $r$, $W_{F50}$ comes out to be the most stable parameter with average correlation of $\bar{r}_{WF50}=0.975$ with the four other linewidths. This is ideal because it is also the linewidth used in our calibration of an isophotal NIR TF relation which is optimal for use in the ZOA \citep{2015MNRAS.447.1618S}

In addition to the preferred linewidth $W_{F50}$, we provide the derived conversion equations between all the five linewidths. These conversion equations are derived using a Bayesian mixture model with more parameters than the data points (see also, \citealt{2010arXiv1008.4686H,astroMLText}). The advantage of using a Bayesian mixture model is two-fold; it is less sensitive to outliers compared to the simple maximum likelihood approach, and it does not exclude them completely as happens with the sigma-clipping procedure. This model was computed with the Python MCMC Hammer {\bf emcee}\footnote{http://dan.iel.fm/emcee/current/} \citep{2013PASP..125..306F}. The 10 conversion formulae are given in Eqs. \ref{eqtocomp}:

\begin{subequations}
\begin{eqnarray}
W_{M50} &=& -0.287(\pm0.086)+1.0089(\pm0.0003) W_{P20},\\
W_{P50} &=& -14.177(\pm0.066)+0.9287(\pm0.0003) W_{P20},\\
W_{P50} &=& -9.053(\pm0.065)+0.9106(\pm0.0003) W_{M50},\\
W_{F50} &=& -27.999(\pm0.181)+1.0474(\pm0.0008) W_{P20},\\
W_{F50} &=& -0.130(\pm0.173)+0.9261(\pm0.0007) W_{M50},\\
W_{F50} &=& 2.554(\pm0.155)+1.0382(\pm0.0007) W_{P50},\\
W_{2P50} &=& -17.293(\pm0.172)+0.9765(\pm0.0009) W_{P20},\\
W_{2P50} &=& -17.293(\pm0.181)+0.9767(\pm0.0009) W_{M50},\\
W_{2P50} &=& -1.400(\pm0.160)+1.0380(\pm0.0009) W_{P50},\\
W_{2P50} &=& -27.280(\pm0.152)+1.0341(\pm0.0008) W_{F50}.
\end{eqnarray}
\label{eqtocomp}
\end{subequations}

Before using any of these linewidths in a TF application, four corrections need to be applied to obtain the more physical rotational velocity, namely: the instrumental resolution correction, the cosmological redshift correction, and the correction for the effect of turbulent motion and inclination. We will apply these four corrections in the forthcoming TF analysis paper (Said et al. in prep).



\subsection{\protect\HI\ Catalog}
The \HI\ catalog and plots of the \HI\ spectra are available electronically. Figure \ref{narrowandhizoa} and Tables \ref{T4} \& \ref{T5} present examples of the \HI\ profiles and catalog. The \HI\ parameters listed in the two catalogs are:

Column (1) - HIZOA ID as reported in the HIZOA survey publications (\citealt{2005AJ....129..220D}; Staveley-Smith et al., 2015; Kraan-Korteweg et al., in prep.).

Columns (2 and 3) - Right Ascension (RA) and Declination (Dec.) in the J2000.0 epoch of the fitted position in HIZOA.

Columns (4 and 5) - Galactic coordinates.

Column (6) - $I$, Integrated \HI\ flux and associated error.

Column (7) - Heliocentric velocity $V_{hel}$ ($cz$) measured as the mid-point of the 50\% of the profile using the $W_{F50}$ algorithm and associated error.

Column (8) - $W_{P20}$, the linewidth at 20\% of the peak flux-rms and associated error.

Column (9) - $W_{M50}$, the linewidth at 50\% of the mean flux and associated error.

Column (10) - $W_{P50}$, the linewidth at 50\% of the peak flux-rms and associated error.

Column (11) - $W_{F50}$, the linewidth at 50\% of the peak flux-rms measured with a polynomial fit to both sides of the profile and associated error.

Column (12) - $W_{2P50}$, the linewidth at 50\% of the peak flux-rms measured at each of the two peaks and associated error.

Column (13) - rms noise.

Column (14) - Signal-to-noise ratio (S/N).\\

\begin{table*}
\begin{turn}{90}
\begin{minipage}{245mm}
\begin{center}
\caption[HIZOA]{\HI\ derived parameters for the good HIZOA profiles (velocity resolution of 27 km s$^{-1}$). The full table is available online.}
\begin{tabular}{ c  c c c c  c c c c c  c c c c}
 \hline
 \hline
\HI\ Name & RA & DEC & $l$ & $b$ & $I$ & $v_{hel}$ & $W_{P20}$ & $W_{M50}$ & $W_{P50}$ & $W_{F50}$ & $W_{2P50}$ & rms & SNR\\
 & \multicolumn{2}{c}{(J2000)} & \multicolumn{2}{c}{[deg]} & Jy km s$^{-1}$  & \multicolumn{6}{c}{km s$^{-1}$} & [mJy] &\\
(1) & (2) & (3) & (4) & (5) & (6)  & (7) & (8) & (9) & (10) & (11) & (12) & (13) & (14)\\
\hline
 J0653-03A &  06 53 21.1 &  -03 53   32 &   216.6146 &    -1.3488 &        8.2 $\pm$    1.0 &    2566.43 $\pm$   2.90 &    290.356 $\pm$   9.42 &     282.72 $\pm$   8.13 &     181.06 $\pm$   4.51 &     190.67 $\pm$   3.19 &     181.06 $\pm$   7.24 &        3.7 &       13.9\\
  J0700-11 &  07 00 58.1 &  -11 47   17 &   224.5099 &    -3.2637 &       18.5 $\pm$    0.7 &    2744.06 $\pm$   1.62 &     449.90 $\pm$  13.19 &     452.45 $\pm$   9.69 &     424.85 $\pm$  10.00 &     426.76 $\pm$   9.25 &     426.80 $\pm$  10.07 &        2.3 &       33.4\\
  J0705-12 &  07 05 39.9 &  -12 59   55 &   226.1123 &    -2.7952 &        5.8 $\pm$    0.7 &    5455.80 $\pm$   2.20 &     222.06 $\pm$  18.49 &     222.99 $\pm$  15.81 &     206.61 $\pm$  10.79 &     209.77 $\pm$   4.30 &     208.88 $\pm$  10.54 &        3.0 &       13.9\\
  J0709-05 &  07 09 34.2 &  -05 24   14 &   219.8091 &     1.5520 &       14.4 $\pm$    1.2 &    1720.60 $\pm$   4.36 &     311.59 $\pm$   1.74 &     305.04 $\pm$   4.79 &     273.74 $\pm$   2.70 &     279.29 $\pm$   2.90 &     273.74 $\pm$   2.79 &        3.8 &       16.5\\
  J0717-08 &  07 17 40.6 &  -08 55   25 &   223.8612 &     1.7025 &        7.0 $\pm$    0.8 &    2460.64 $\pm$   4.45 &     201.84 $\pm$   4.59 &     205.05 $\pm$   3.62 &     178.40 $\pm$   4.25 &     187.15 $\pm$  17.97 &     178.40 $\pm$   4.25 &        3.4 &       15.0\\
  J0722-09 &  07 22 47.2 &  -09 01   54 &   224.5481 &     2.7670 &        9.0 $\pm$    1.2 &    3338.34 $\pm$   4.83 &     219.51 $\pm$  22.20 &     231.46 $\pm$  21.04 &     187.63 $\pm$  10.82 &     181.50 $\pm$  20.73 &     187.63 $\pm$  10.82 &        4.5 &       13.1\\
  J0724-09 &  07 24 57.5 &  -09 39   18 &   225.3511 &     2.9460 &       56.7 $\pm$    0.7 &    2438.02 $\pm$   2.32 &     203.50 $\pm$  13.66 &     211.72 $\pm$  11.86 &     178.45 $\pm$   6.33 &     220.53 $\pm$  17.65 &     178.45 $\pm$   6.33 &        2.7 &      130.5\\
 J0725-24A &  07 25 16.4 &  -24 28   19 &   238.4486 &    -3.9987 &       57.5 $\pm$    0.8 &     796.53 $\pm$   0.36 &     195.95 $\pm$   0.43 &     257.60 $\pm$   5.64 &     131.04 $\pm$   3.98 &     135.94 $\pm$   0.16 &     131.04 $\pm$   3.98 &        3.0 &      125.2\\
 J0725-24B &  07 25 16.5 &  -24 57   20 &   238.8765 &    -4.2258 &       10.8 $\pm$    0.8 &    2761.76 $\pm$   2.12 &     266.37 $\pm$  16.29 &     267.44 $\pm$  14.38 &     244.33 $\pm$   5.19 &     245.66 $\pm$   7.51 &     246.71 $\pm$   6.29 &        2.9 &       19.6\\
  J0727-23 &  07 27 31.3 &  -23 57   48 &   238.2420 &    -3.3068 &       18.7 $\pm$    0.9 &    4391.43 $\pm$   1.00 &     442.93 $\pm$   9.04 &     445.06 $\pm$   6.70 &     417.02 $\pm$   6.70 &     420.80 $\pm$   8.78 &     420.06 $\pm$   9.18 &        2.9 &       26.2\\

\hline
\end{tabular}
\label{T4}
\end{center}

\begin{center}
\caption[Narrow]{\HI\ derived parameters for the narrow-band profiles  (velocity resolution of 3.3 km s$^{-1}$). The full table is available online.}
\begin{tabular}{ c  c c c c  c c c c c  c c c c}
 \hline
 \hline
\HI\ Name & RA & DEC & $l$ & $b$ & $I$ & $v_{hel}$ & $W_{P20}$ & $W_{M50}$ & $W_{P50}$ & $W_{F50}$ & $W_{2P50}$ & rms & SNR\\
 & \multicolumn{2}{c}{(J2000)} & \multicolumn{2}{c}{[deg]} & Jy km s$^{-1}$  & \multicolumn{6}{c}{km s$^{-1}$} & [mJy] &\\
 (1) & (2) & (3) & (4) & (5) & (6)  & (7) & (8) & (9) & (10) & (11) & (12) & (13) & (14)\\
\hline
  J0744-13 &  07 44 29.5 &  -13 03   58 &   230.6608 &     5.4984 &       21.9 $\pm$    0.7 &    2304.82 $\pm$   0.44 &     128.96 $\pm$   3.01 &     125.94 $\pm$   0.58 &     111.39 $\pm$   2.30 &     115.93 $\pm$   1.62 &     114.21 $\pm$   1.80 &       10.3 &      23.14\\
  J0744-25 &  07 44 26.4 &  -25 59   20 &   241.8791 &    -0.9424 &        2.1 $\pm$    0.3 &    4048.62 $\pm$  12.66 &     182.10 $\pm$   2.28 &     174.57 $\pm$   6.76 &     131.00 $\pm$  45.66 &     170.17 $\pm$  12.55 &     150.57 $\pm$   2.78 &        4.4 &       6.44\\
  J0744-26 &  07 44 23.4 &  -26 01   48 &   241.9092 &    -0.9725 &        8.8 $\pm$    0.5 &    2728.33 $\pm$   0.74 &     318.23 $\pm$   1.37 &     309.14 $\pm$   1.92 &     288.18 $\pm$   9.41 &     285.73 $\pm$  12.41 &     268.52 $\pm$  34.02 &        5.4 &       9.02\\
  J0744-27 &  07 44 55.4 &  -27 21   10 &   243.1148 &    -1.5301 &        3.5 $\pm$    0.4 &    7272.13 $\pm$   7.54 &     313.30 $\pm$   2.10 &     314.20 $\pm$   3.41 &     252.10 $\pm$  51.65 &     267.42 $\pm$  40.15 &     273.47 $\pm$  12.98 &        5.6 &       6.17\\
  J0747-21 &  07 47 19.7 &  -21 37    9 &   238.4298 &     1.8211 &        2.9 $\pm$    0.4 &    6969.20 $\pm$   3.39 &     123.33 $\pm$   9.66 &     120.45 $\pm$   0.32 &     106.60 $\pm$   9.57 &     108.67 $\pm$  16.82 &     107.35 $\pm$  31.65 &        6.0 &       7.32\\
 J0748-25B &  07 48 33.6 &  -25 14   35 &   241.7011 &     0.2355 &        5.2 $\pm$    0.6 &    6813.39 $\pm$   6.13 &     387.72 $\pm$  11.11 &     387.31 $\pm$   8.30 &     359.04 $\pm$  26.94 &     381.84 $\pm$  24.89 &     375.88 $\pm$  14.65 &        6.5 &       4.55\\
  J0749-21 &  07 49 21.4 &  -21 53    6 &   238.8975 &     2.0935 &        2.9 $\pm$    0.4 &    2338.38 $\pm$   0.33 &     192.44 $\pm$   0.08 &     182.80 $\pm$   2.00 &     165.54 $\pm$  15.94 &     179.23 $\pm$   6.77 &     174.32 $\pm$  11.21 &        5.7 &       5.59\\
 J0749-26B &  07 49 21.0 &  -26 12   17 &   242.6205 &    -0.0983 &       13.0 $\pm$    0.5 &    2490.82 $\pm$   1.42 &     328.04 $\pm$  10.59 &     313.38 $\pm$   0.07 &     279.72 $\pm$   8.94 &     296.52 $\pm$   4.76 &     288.81 $\pm$   2.63 &        4.4 &      13.74\\
  J0750-32 &  07 50 39.7 &  -32 49   19 &   248.4695 &    -3.2120 &        5.0 $\pm$    0.4 &    5130.45 $\pm$   0.36 &     329.91 $\pm$   6.04 &     320.56 $\pm$   0.33 &     290.53 $\pm$   7.32 &     307.97 $\pm$  13.66 &     303.12 $\pm$   6.95 &        4.3 &       7.68\\
  J0751-37 &  07 51 28.0 &  -37 13    4 &   252.3550 &    -5.2946 &       12.4 $\pm$    0.5 &    2805.47 $\pm$   0.28 &     101.69 $\pm$   2.71 &      99.19 $\pm$   2.60 &      79.94 $\pm$   3.18 &      85.73 $\pm$   0.30 &      85.97 $\pm$   1.49 &        8.4 &      18.78\\

\hline
\end{tabular}
\label{T5}
\end{center}

\end{minipage}
\end{turn}
\end{table*}

\subsection{Characteristics of the Current Data}
\begin{figure*}
\begin{center}
\includegraphics[scale=0.5,angle=270]{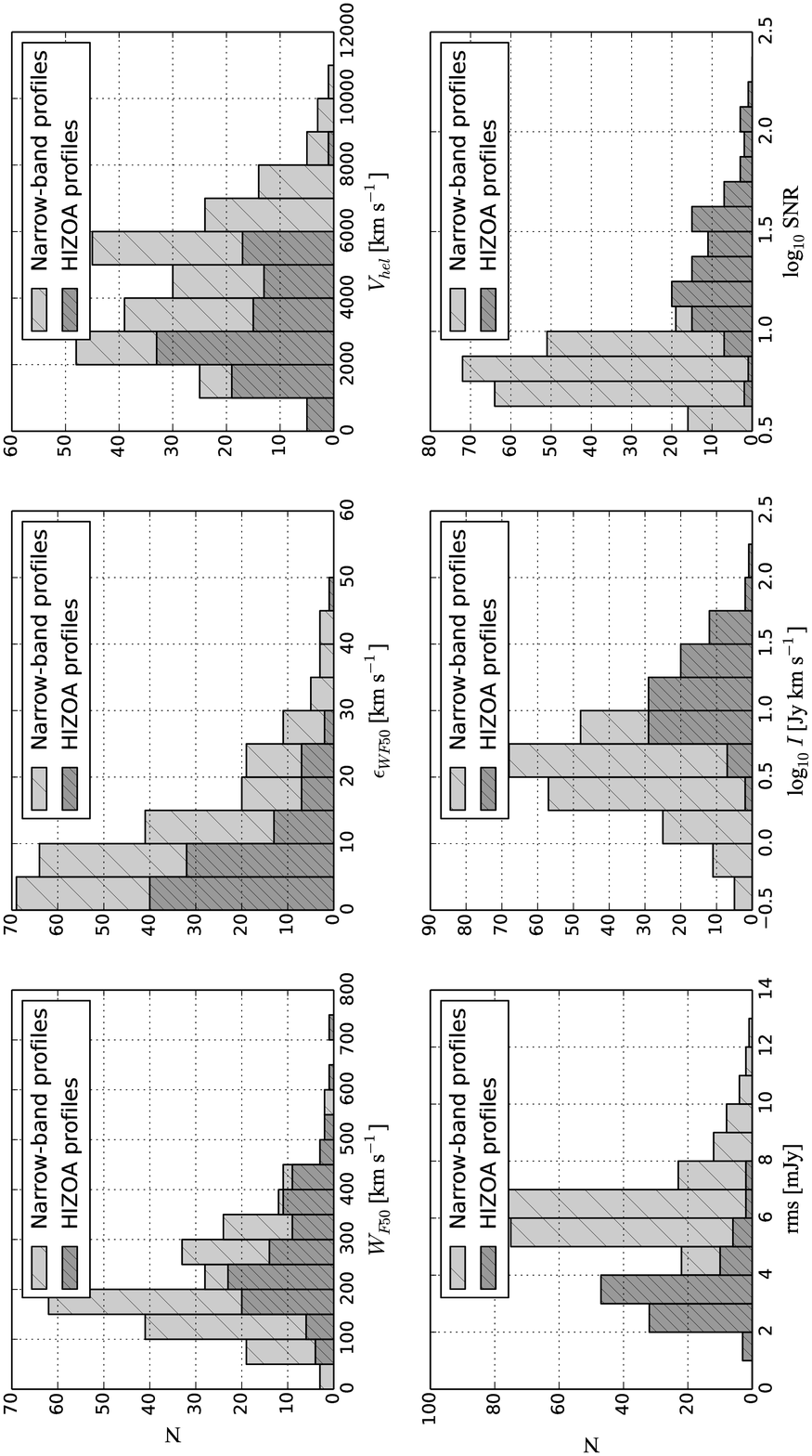}
\caption{Histograms of the \protect\HI\ parameters for the whole TF sample divided into two sub-samples (238 narrow band profiles and 104 HIZOA profiles); the linewidth at 50\% of the peak flux$-$rms measured after fitting polynomials to both sides of the profile $W_{F50}$, the error on the linewidth $\epsilon_{WF50}$, the heliocentric velocity $V_{hel}$, the rms noise, logarithmic integrated line flux and logarithmic signal-to-noise ratio.}
\label{Characteristics}
\end{center}
\end{figure*}

An overview of the \HI\ parameters (divided into two sub-samples) is given in Fig. \ref{Characteristics}, which shows histograms of the linewidth at 50\% of the peak flux$-$rms measured after fitting polynomials to both sides of the profile $W_{F50}$, the error on the linewidth $\epsilon_{WF50}$, the heliocentric velocity $V_{hel}$, the rms noise, the logarithmic integrated line flux and the logarithmic signal-to-noise ratio.

The distribution of galaxies with linewidths $W_{F50} > 150$ km s$^{-1}$ is similar to the distribution of the overall HIZOA catalogue (see also Fig. 2 in Staveley-Smith et al., 2015). The fraction of low-linewidth galaxies in the TF sample is significantly lower. That is to be expected because these profiles generally originate from dwarf galaxies which often have no clear disks, hence uncertain inclinations, and low $S/N$. They are also less likely to have a NIR counterpart. The scatter in the TF relation for these low-linewidth galaxies is much higher and they are preferentially excluded from TF applications.

The distribution of the associated linewidth error $\epsilon_{WF50}$ in Fig. \ref{Characteristics} shows that 95\% and 84\% of our sample have errors less than 30 km s$^{-1}$ and 20 km s$^{-1}$, respectively. 

The distribution of the heliocentric velocity $V_{hel}$ shows the peaks that are indicative of the Puppis and GA overdensities, but they are not quite as pronounced as in the HIZOA data. There is a clear drop in the number of detected galaxies for reliable TF analysis beyond 7000 km s$^{-1}$, compared to the full HIZOA data set. Only 24 galaxies (7\%) of the sample have $V_{hel}$ > 7000 km s$^{-1}$. Galaxies beyond 7000 km s$^{-1}$ on average have lower $S/N$ ratios given their large distances.

The average rms noise for our sample is 5.5 mJy and the majority (92\%) have rms < 8.0 mJy. A histogram of the logarithmic value of the integrated line flux is also shown in Fig. \ref{Characteristics}. The final histogram in Fig. \ref{Characteristics} shows the distribution of the logarithmic value of the signal-to-noise ratio. The average value of the signal-to-noise ratio of our sample is 14.7 with a median value of 7.9, which is adequate for the TF analysis. 

In Fig. \ref{Characteristics}, the 104 HIZOA profiles have higher S/N ratio than the new narrow band observations. This is expected because 
the narrow-band observing list contains only galaxies with weaker HIZOA profiles of insufficient S/N ratio for the TF work.

\subsection{Comparison with published \protect\HI\ data}
In this section we provide an extensive comparison of the newly measured linewidths and the published linewidths. We used the two published parts of the HIZOA survey (HIZOA-S; Staveley-Smith et al., 2015 and HIZOA-N; \citealt{2005AJ....129..220D}). After excluding galaxies from the GB extension, the number of galaxies used in this comparison becomes 307 galaxies.

\begin{figure}
\begin{center}
\includegraphics[scale=0.42,angle=270]{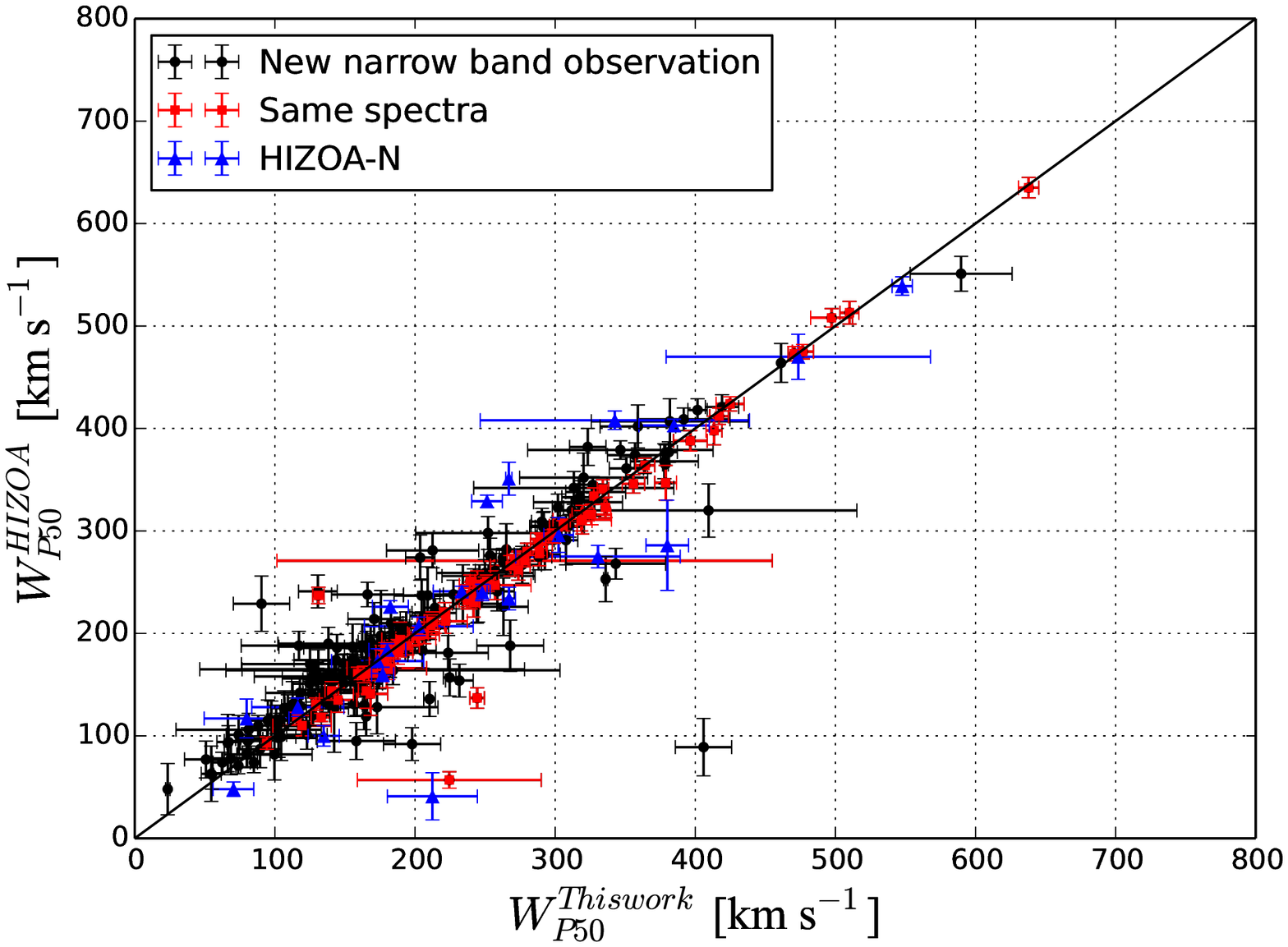}
\caption{A comparison between $W_{P50}$ measured in this work using a modified version of the GBTIDL function AWV and $W_{P50}$ measured in HIZOA surveys using MIRIAD task MBSPECT. The solid line shows the one-to-one relation.}
\label{notable_offsets}
\end{center}
\end{figure}

Figure \ref{notable_offsets} shows a comparison between linewidths at 50\% of the peak flux measured in this work against its counterpart measurements in HIZOA surveys.
In Fig. \ref{notable_offsets}, we divided the whole sample into three sub-sample: (1) the new narrow band observations, (2) HIZOA profiles (same spectra) and (3) The northern extinction profiles (HIZOA-N). Good agreement for most of the galaxies are shown clearly in this figure. Measurements from the same spectra, shown as red squares, are almost identical. Our new measurements are slightly different from the HIZOA-N (blue triangles). In Table \ref{offsets} we list all galaxies with offsets larger than 50 km s$^{-1}$.
\begin{table}
\begin{center}
\caption[List of galaxies with offset larger than 50 km s$^{-1}$]{List of galaxies with offset larger than 50 km s$^{-1}$}
\begin{tabular}{l l l r}
\hline
\hline
\HI\ Name  &   $W_{P50}^{This\/work}$ & $W_{P50}^{HIZOA}$ & offset\\
\hline
J0623+14 & $380\pm15$ & $286\pm44$ & $94$\\
J0635+02 & $212\pm32$ & $41\pm23$ & $171$\\
J0635+11 & $342\pm96$ & $408\pm9$ & $-65$\\
J0635+14A & $330\pm58$ & $275\pm11$ & $55$\\
J0653+07 & $251\pm10$ &  $329\pm6$ & $-77$\\
J0704-13 & $197\pm20$ & $92\pm18$ & $105$\\
J0722-05 & $130\pm13$ & $241\pm16$ & $-110$\\
J0730-25 & $336\pm1$ & $253\pm23$ & $83$\\
J0752-29 & $166\pm25$ & $238\pm14$ & $-71$\\
J0753-22 & $231\pm9$ & $154\pm18$ & $77$\\
J0808-35 & $409\pm50$ & $320\pm26$ & $89$\\
J0821-39 & $267\pm23$ & $188\pm25$ & $79$\\
J0858-45A & $343\pm35$ & $268\pm15$ & $75$\\
J0859-52 & $210\pm4$ & $136\pm18$ & $74$\\
J1012-62 & $90\pm20$ & $229\pm27$ & $-138$\\
J1045-64 & $323\pm13$ & $382\pm18$ & $-58$\\
J1052-64 & $158\pm27$ & $95\pm18$ & $63$\\
J1125-60 & $224\pm4$ & $157\pm19$ & $67$\\
J1149-64 & $117\pm41$ & $188\pm14$ & $-70$\\
J1339-57 & $212\pm32$ & $281\pm17$ & $-68$\\
J1419-57 & $138\pm35$ & $190\pm16$ & $-51$\\
J1624-45A & $203\pm10$ & $274\pm22$ & $-70$\\
J1625-55 & $405\pm20$ & $89\pm28$ & $316$\\
J1929+11 & $266\pm2$ & $351\pm16$ & $-84$\\
J0858-39 & $224\pm65$ & $57\pm8$ & $167$\\
\hline
\end{tabular}
\label{offsets}
\end{center}
\end{table} 
Table \ref{offsets} also shows the newly measured, the published linewidth with the associated error for each measurement and the offset for each galaxy.

The main reason for these offsets is the difference in the velocity resolution between the narrow band (3.3 km s$^{-1}$) and the HIZOA (27 km s$^{-1}$) data. This is shown clearly in the most notable offset (316 km s$^{-1}$, for J1625$-$55). This galaxy is measured in the narrow band observations to have $W_{P50} = 405$ km s$^{-1}$, which agrees within the uncertainty in the measurement of the same galaxy in the HICAT catalog \citep{2004MNRAS.350.1195M} which was $W_{P50} = 453$ km s$^{-1}$. In contrast, the same galaxy was measured in the HIZOA-S catalog to have $W_{P50}$ of 89 km s$^{-1}$ because one of the two peaks dominate their signal. This can be further explained by looking at the $W_{P20}$ in the three surveys (narrow-band, HICAT and HIZOA-S) which all agree within the uncertainty. All other notable offsets listed in Table \ref{offsets} can be explained in a similar manner.

\begin{figure*}
\begin{center}
\begin{tabular}{cc}
\multirow{2}{*}{\subfloat[]{\includegraphics[width=3.5in,height=4.4in,angle=270]{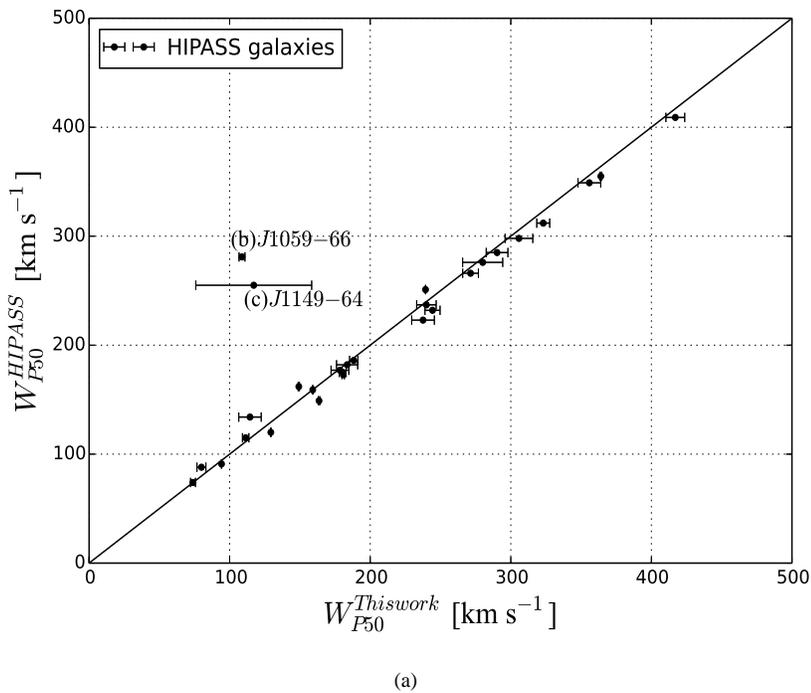}}}
   & \subfloat[]{\includegraphics[width=1.6in,height=2.3in,angle=270]{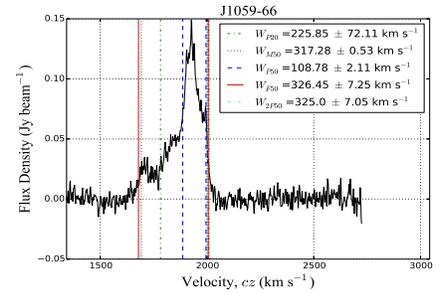}} \\
   & \subfloat[]{\includegraphics[width=1.6in,height=2.3in,angle=270]{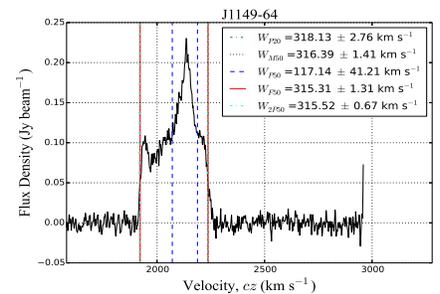}} \\
\end{tabular}
\caption[A comparison between $W_{P50}$ measured in this work using a modified version of the GBTIDL function AWV and $W_{P50}$ measured in HIPASS survey using MIRIAD task MBSPECT. The solid line shows the one-to-one relation.]{A comparison between $W_{P50}$ measured in this work using a modified version of the GBTIDL function AWV and $W_{P50}$ measured in HIPASS survey using MIRIAD task MBSPECT. The solid line shows the one-to-one relation.}
\label{hipass_comp}
\end{center}
\end{figure*}

Another comparison with the HIPASS catalog \citep{2004AJ....128...16K} is shown in Fig. \ref{hipass_comp}.

The left panel of this plot shows a comparison between $W_{P50}$ measured in this work on the x-axis and $W_{P50}$ measured in HIPASS survey on the y-axis for the 28 galaxies in common. As expected, a perfect agreement between both measurements are shown because these galaxies are among the brightest galaxies in the HIPASS survey. Two galaxies, J1059-66 \& J1149-64, both show a notable offset. They are identified by name on the left panel of the figure itself. The profiles of these two galaxies are
displayed in the right panel of the plot. They demonstrate quite well why there is such a large difference:  in both cases one of the two peaks is more prominent than the other peak in these lob-sided profiles.

\section{Summary}
In this paper, we present \HI\ observations for 394 inclined spiral galaxies in the southern ZOA. Parameters were determined in a systematic way for applications of the TF relation in the ZOA. New observations for 290 galaxies in the sample were conducted in 2010 and 2015 with the 7 inner beams of the multibeam receiver  on the Parkes 64 m Radio Telescope. Three-channel Hanning smoothing was applied to the spectra, resulting in a velocity resolution of 3.3 km s$^{-1}$ for our newly observed sample. The additional 104 galaxies were measured directly from the existing HIZOA data. The final sample contains 342 galaxies with adequate signal-to-noise ($S/N > 5$) and an \HI\ profiles suitable for TF analysis. The average value of the signal-to-noise ratio of the final sample is 14.7 with a median value of 7.9. A modified version of the {\bf GBTIDL} function {\bf AWV} was used to measure the systemic velocities and velocity linewidths based on five different methods. Good  agreement was found, within the uncertainty, between $W_{P20}$, the linewidth at 20\% of the peak flux-rms and $W_{M50}$, the linewidth at 50\% of the mean flux. A larger offset was found between both of these linewidth values and $W_{F50}$, the linewidth at 50\% of the peak flux-rms measured with a polynomial fit to both sides of the profile. Also, $W_{F50}$, the linewidth at 50\% of the peak flux-rms and $W_{2P50}$, the linewidth at 50\% of the peak flux-rms measured at each of the two peaks seem to agree within the uncertainty. The Pearson's sample correlation coefficient was used to quantify the correlation between all derived linewidths.  Conversion equations between these five linewidths were derived using a Bayesian mixture model to avoid any bias toward the outliers. 

In a forthcoming paper, Said et al. in prep., we will present the accompanying final high-quality NIR $J$, $H$, and $K_s$-band photometry that has been obtained for all galaxies in the HIZOA surveys. In that paper we will also discuss the completeness of the NIR catalog in the ZOA and the reliability of the photometry with regard to the determinations of peculiar velocities from the NIR TF  application.

The \HI\ data presented in this paper, the NIR data (Said et al. in prep.) and the newly calibrated TF relation \citep{2015MNRAS.447.1618S} will be used to measure the flow fields in the ZOA. This will be the first measurement of the flow fields in the ZOA and should improve the assessment of the actual overdensity of the so far hidden large-scale structures, in particular for the GA region. This work should also be regarded as a precursor to the forthcoming Widefield ASKAP L-band Legacy All-sky Blind surveY (WALLABY)\footnote{http://www.atnf.csiro.au/research/WALLABY/proposal.html} and its sister in the northern hemisphere, the Westerbork Northern Sky HI Survey (WNSHS)\footnote{http://www.astron.nl/~jozsa/wnshs/} which will provide an HI survey of the whole sky \citep{2012MNRAS.426.3385D}. This work can be regarded as a pilot project as well to the determination of flow fields as close as possible  to the Galactic Plane, given the forthcoming deep NIR photometry from surveys such as UKIDSS and VISTA \citep{2004SPIE.5493..401E,2006MNRAS.367..454H,2008MNRAS.391..136L}.

\section*{Acknowledgments}
This work is based upon research supported by the Science Faculty at University of Cape Town (UCT), the South African National Research Foundation and the Department of Science and Technology. We acknowledge the HIZOA survey team for early access to the data. And foremost, we thank the CSIRO staff Stacy Mader for the valuable support provided during the observations. We also thank an anonymous referee for various useful suggestions. Part of this research was conducted by the Australian Research Council Centre of Excellence for All-sky Astrophysics (CAASTRO), through project number CE110001020.

\bibliographystyle{mn2e.bst}
\bibliography{774}

\label{lastpage}

\bsp

\end{document}